# DEGRESSIVE PROPORTIONALITY IN THE EUROPEAN PARLIAMENT

Katarzyna Cegiełka

**Abstract.** The division of mandates to the European Parliament has posed difficulties since the beginning of its operation. The introduction of the degressive proportionality principle to legal acts intensified further discussion on this subject. The researchers propose different solutions in the form of algorithms or functions with which it is possible to determine the composition of the European Parliament.

## 1. Introduction

The division of goods is one of the oldest problems of any kind of community. It concerns, *inter alia*, the separation of objects that cannot be divided into smaller parts. This situation occurs in the case of the distribution of parliamentary mandates. Over hundreds of years various methods of determining the composition of the units representing a particular community have been developed. These include the known method of Hamilton and the divisor methods of Jefferson, Adams and Webster (Cegiełka et al., 2010a). However, they are based on proportional distribution of the use of which is not possible in the case of the European Parliament. This results from the large diversity of the Member States in terms of the number of citizens, based on which seats are allocated. For this reason, the composition of the European Parliament is determined by the principle of degressive proportionality (Cegiełka et al., 2010b).

## 2. Degressive proportionality in legal acts

Degressive proportionality, introduced in order to standardize the rules for distributing seats in the European Parliament was enshrined in Article 1, Point 15 of the Treaty of Lisbon: "The European Parliament shall be composed of representatives of the Union's citizens. They shall not exceed seven hundred and fifty in number, plus the President. Representation of citizens shall be degressively proportional, with a minimum threshold of six members per Member State. No Member State shall be allocated more than ninety-six seats" (*The Treaty of Lisbon*, 2010).

This legacy points out the total number of mandates and the minimum and maximum number of seats for each country. It does not, however, include the characteristics of degressive proportionality. This lack was supplemented in the

---

**Katarzyna Cegiełka**
Department of Mathematics, Wrocław University of Economics, Komandorska Street 118/120, 53-345 Wrocław, Poland.
E-mail: katarzyna.cegielka@ue.wroc.pl

Report of the Committee on Constitutional Affairs and the European Parliament Resolution. The rules included in them specify how to apply the principle introduced in the Lisbon Treaty. Analysis of the contents of these rules allows distinguishing two conditions with which a degressively proportional division must comply. According to them, members from countries with a higher population represent a greater number of citizens and do not have fewer seats than the less populated countries (Cegiełka et al., 2010b).

Denoting as $n$ – number of countries, $l_i$ – population of the country and $m_i$ – the number of mandates of the country, all the conditions can be saved as follows (Cegiełka, 2010):

$W1.$ $\sum_{i=1}^{n} m_i = 751$, $6 \leq m_i \leq 96$,

$W2.$ $l_1 < l_2 < \cdots < l_n \Rightarrow m_1 \leq m_2 \leq \cdots \leq m_n$,

$W3.$ $l_1 < l_2 < \cdots < l_n \Rightarrow \dfrac{l_1}{m_1} < \dfrac{l_2}{m_2} < \cdots < \dfrac{l_n}{m_n}$.

The first of these is contained in the Treaty itself – the total number of MPs does not exceed 751, but no country may receive less than 6 and more than 96 seats. Others stem from the rules included in additional parliamentary documents. The second says that states with smaller populations may not receive more seats than states with larger populations. The third indicates that MEPs from countries with larger populations represent a greater number of citizens than MEPs from countries with a lower population.[2]

### 3. Definitions of degressive proportionality

#### 3.1. "Strong" degressive proportionality

Differences in the population of the Member States meant that it was decided to determine a "just, comprehensible and lasting system for the distribution of seats" (Lamassoure, Severin, 2007).

The result of this work was the introduction of the principle of degressive proportionality. The Lisbon Treaty, in which it was written, entered into force in December 2009, diversification of the EU countries had been in place since its inception. Within several dozen years of the functioning of the European Parliament, there were attempts to formulate the method of selecting its composition. One of them is the algorithm set at the meeting of the Council of Europe in 1992. It can be shown in 4 points (Lamassoure, Severin, 2007):

1) Each state receives 6 seats.

2) States with a population of 1 to 25 million receive a mandate for every 500 thousand citizens.

3) States with a population of 25 do 60 million receive a mandate for every million citizens.

---

[2] For a detailed mathematical analysis of the degressive proportionality, see Florek (2011).

4) States with a population exceeding 60 million receive a mandate for every 2 million citizens.

This pattern was never strictly applied. However, it indicates some sort of division – to assign successive mandates to an increasing number of represented citizens. It turns out that this type of division is always consistent with the principle of degressive proportionality.

Degressively proportional division will, therefore, also be the division that meets the following conditions:

$$W'1. \quad \sum_{i=1}^{n} m_i = 751, \ 6 \leq m_i \leq 96,$$

$$W'2. \quad l_1 < l_2 < \cdots < l_n \Rightarrow m_1 \leq m_2 \leq \cdots \leq m_n,$$

$$W'3. \quad \frac{\beta}{\beta} < \frac{l_1}{m_1} < \frac{\alpha_2}{\mu_2} < \cdots < \frac{\alpha_i}{\mu_i} < \frac{\alpha_{i+1}}{\mu_{i+1}} < \cdots < \frac{\alpha_n}{\mu_n},$$

where $\alpha_i = l_i - l_{i-j-1}, \mu_i = m_i - m_{i-j-1}$, when $m_{i-j-1} < m_{i-j} = m_{i-j+1} = \cdots = m_{i-1} = m_i$, and $\alpha_i = l_i - \beta$, $\mu_i = m_i - \beta$, when $m_1 = m_{i-j} = m_{i-j+1} = \cdots = m_{i-1} = m_i$ for $i = \{2,3,\ldots,n\}, j = i - 2$, and $0 < \beta < m_1 < l_1$. We also define $m_1 = \mu_1$ and $l_1 = \alpha_1$.

**Theorem:** Each division meeting the conditions $W'1, W', W'3$ also meets conditions $1, W2, W3$.

PROOF. Conditions $W'1$ and $W'2$ are respectively equal to the conditions $W1$ and $W2$; therefore, it suffices to show:

$$\forall_i \frac{\alpha_i}{\mu_i} < \frac{\alpha_{i+1}}{\mu_{i+1}} \Rightarrow \frac{l_i}{m_i} < \frac{l_{i+1}}{m_{i+1}}.$$

According to the definition of the quotient $\frac{\alpha_i}{\mu_i}$, inequality $\frac{\alpha_i}{\mu_i} < \frac{\alpha_{i+1}}{\mu_{i+1}}$ is equivalent to one of the following inequalities:

$$\frac{l_i - l_{i-1}}{m_i - m_{i-1}} < \frac{l_{i+k} - l_{i-1}}{m_{i+k} - m_{i-1}}; \ m_{i-1} < m_i = m_{i+1} \ldots = m_{i+k}, \quad (1)$$

$$\frac{l_i - l_{i-1}}{m_i - m_{i-1}} < \frac{l_{i+k} - l_{i+k-1}}{m_{i+k} - m_{i+k-1}}; \ m_{i-1} < m_i = m_{i+1} = \cdots = m_{i+k-1} < m_{i+k}. \quad (2)$$

If $m_1 = m_2$ expressions $l_{i-1}, m_{i-1}$ from inequalities (1) and (2) are replaced by $\beta$ and the proof is analogous.

For $i = \{2,\ldots,n-1\}, \ k = \{1,2,\ldots,n-2\}, k \leq n - i$ and inequality (1) we have:

$$\frac{\alpha_i}{\mu_i} < \frac{\alpha_{i+k}}{\mu_{i+k}} \Leftrightarrow \frac{l_i - l_{i-1}}{m_i - m_{i-1}} < \frac{l_{i+k} - l_{i-1}}{m_{i+k} - m_{i-1}}.$$

By assumption $m_{i+k} = m_i$, inequality (1) is then equivalent to

$$\frac{l_i}{m_{i+k} - m_{i-1}} < \frac{l_{i+k}}{m_{i+k} - m_{i-1}} \Leftrightarrow \frac{l_i}{m_i} < \frac{l_{i+k}}{m_{i+k}}.$$

Which in particular for $k = 1$ implies $\frac{\alpha_i}{\mu_i} < \frac{\alpha_{i+1}}{\mu_{i+1}} \Leftrightarrow \frac{l_i}{m_i} < \frac{l_{i+1}}{m_{i+1}}$.

For the inequality (2) and by definition the quotient $\frac{\alpha_i}{\mu_i}$ we obtain:

$$\frac{\alpha_i}{\mu_i} < \frac{\alpha_{i+k}}{\mu_{i+k}} \Leftrightarrow \frac{l_i - l_{i-1}}{m_i - m_{i-1}} < \frac{l_{i+k} - l_{i+k-1}}{m_{i+k} - m_{i+k-1}}.$$

Inequality $\frac{l_i}{m_i} < \frac{\alpha_{i+k}}{\mu_{i+k}}$ implies $\frac{l_i}{m_i} < \frac{l_{i+k}}{m_{i+k}}$, which results from the definition of the quotient $\frac{\alpha_i}{\mu_i}$.

Therefore, it suffices to show that $\frac{l_i}{m_i} < \frac{\alpha_{i+k}}{\mu_{i+k}}$. Taking into account the definition of $\alpha_1$ and $\mu_1$ and condition $W'3$ saying that $\frac{l_1}{m_1} < \frac{\alpha_2}{\mu_2} < \cdots < \frac{\alpha_i}{\mu_i} < \cdots < \frac{\alpha_n}{\mu_n}$ we have $\frac{l_i}{m_i} < \frac{\alpha_{i+k}}{\mu_{i+k}}$. Thus

$$\frac{\alpha_i}{\mu_i} < \frac{\alpha_{i+1}}{\mu_{i+1}} \Rightarrow \frac{l_i}{m_i} < \frac{l_{i+1}}{m_{i+1}}.$$

We have shown that each division that satisfies the conditions $W'1, W', W'3$ also satisfies conditions $W1, W2, W3$. This means that the determination of the composition of the European Parliament which consist in allocating more seats to Members representing a growing number of citizens always generates a degressively proportional division. The construction of division that satisfies the conditions of a "strong" degressive proportionality is much more difficult than the determination of the composition satisfying the conditions enclosed in the European Parliament Resolution. This problem is exacerbated with the increasing number of Member States. With the current number of 27 countries, the distribution already poses difficulties that satisfy conditions $W1, W2, W3$. Therefore, researchers have proposed modifying the conditions of degressive proportionality.

### 3.2. "Weakened" degressive proportionality

In February 2011 at the Committee on Constitutional Affairs meeting a group of mathematicians led by Professor Geoffrey Grimmett presented a proposal to standardize the composition of the European Parliament – the so-called Cambridge Compromise. The scientists proposed the method "base+prop", where each State receives a certain number of seats ("base") and then the remaining number of seats is divided by one of the classic methods of proportional allocation ("prop"). They concluded that the best choice is the base equal to five mandates and division of Adams divisor method (assuming rounding fractions up to the nearest whole integer) so that each Member receives a minimum 6 seats as guaranteed in the Treaty of Lisbon. The authors in their deliberations went even further. They considered that – apart from the introduction of an algorithm developed by them – there also should be a change in the definition of degressive proportionality proposed by A. Lamassoure and A. Severin in the Report of the Committee on Constitutional Affairs on the composition of the European Parliament from 2007: "[The European Parliament] considers that the principle of degressive proportionality means that the ratio between the population and the number of seats of each Member State must vary in relation to their respective populations in such a way that each Member from a more populous Member State represents more citizens than each Member from a less populous Member State and conversely, but also that no less populous Member State has more seats than a more populous Member State" (Lamassoure, Severin, 2007).[3]

The authors of the Compromise proposed the following changes: "[The European Parliament] considers that the principle of degressive proportionality means that the ratio between the population and the number of seats of each Member State **before rounding to whole numbers** must vary in relation to their respective populations in such a way that each Member from a more populous Member State represents more citizens than each Member from a less populous Member State and conversely, but also that no less populous Member State has more seats than a more populous Member State" (Grimmett, 2011).

Professor Grimmett also refers to the theorem which states that with this definition the division will always be degressively proportional (*On the apportionment…*). Such a guarantee, however, occurs only in the case of the composition based on a proven algorithm. It is clear that the algorithm or function that assigns the number of seats depending on the population in a degressively proportional manner will return a degressively proportional distribution if the results are not changed (rounded).[4] The "weakness" of the definition, however, brought, beyond the facilitation of accounts to the authors, no solution. Members declare that "the ideal alternative would be to agree on an undisputed mathematical formula of "degressive proportionality" that would ensure a solution not only for the present revision but for future enlargements or modifications due to demographic changes" (Lamassoure, Severin, 2007). So far, however, they have not accepted any of the developed solutions.[5] The only way of selecting the composition of the European Parliament remains then tedious negotiations, as held so far. In this case, MPs shall determine the total number of seats; therefore, a modification of the Lamassoure and Severin's definition leads to nowhere. The introduction of changes without a doubt would, however, simplify the work of the authors of the various functions and algorithms.

---

[3] Which was written down in conditions W2 and W3.

[4] The authors of the various proposals for the function of separating the mandates are well aware that it is the question of the integer number that "spoils" degressive proportionality.

### 4. Conclusions

The European Parliament currently consists of representatives of the citizens of 27 countries, whose populations are characterized by a large dispersion. This leads to the need of seeking the methods of allocating the seats which are not based on proportional methods. According to the Lisbon Treaty, they should, however, fulfill the conditions of degressive proportionality. Scientists have so far offered various solutions in line with the assumptions. However, MPs have not taken any of them. In addition, it appears that the interpretation of degressive proportionality largely depends on the interpreter. A multitude of unknowns and the lack of a determined position of MEPs means that the problem of unification of the procedures for selecting the composition of the European Parliament still remains unsolved.

### Literature

Cegiełka K. (2010). *Distribution of seats in the European Parliament in accordance with the principle of degressive proportionality*. Mathematical Economics 6(13).

Cegiełka K., Dniestrzański P., Łyko J., Misztal A. (2010a). *Division of seats in the European Parliament*. Journal for Perspectives of Economic Political and Social Integration. Vol. XVI.

Cegiełka K., Dniestrzański P., Łyko J., Misztal A. (2010b). *Demographic changes and principles of the fair division*. International Journal Of Social Sciences and Humanity Studies. Vol. 2. No. 2.

Cegiełka K., Dniestrzański P., Łyko J., Misztal A. (2011). *Degressive proportionality in the context of the composition of the European Parliament*. Economics 2(14).

Florek J. (2011). *Allocation of Seats in the European Parliament and a Degressive Proportionality*. http://arxiv.org/abs/1104.3075.

Grimmett G.R. (2011). *European apportionment via the Cambridge Compromise*. Mathematical Social Sciences. http://www.statslab.cam.ac.uk/~grg/papers/USep3.pdf. Accessed: 08.09.2011.

Lamassoure A., Severin A. (2007). *Report on the Composition of the European Parliament*. A6-0351/2007.

*On the Apportionment of the Seats in the European Parliament: A Report by Mathematicians*. Video from the debate.

http://www.europarl.europa.eu/wps-europarl-internet/frd/vod/player;jsessionid=888E10A

---

[5] Members rejected, among others, V. Ramirez-Gonzalez's "Parabolic method" and have not made so far the Cambridge Compromise.

E941B2ACA1B7152FA7A0C038D?category=COMMITTEE&eventCode=201102
07-1600-COMMITTEE-
FCO&format=wmv&byLeftMenu=researchcommittee&
language=en#anchor1. Accessed: 08.09.2011.

*The Treaty of Lisbon* (2010).
http://eur-
lex.europa.eu/LexUriServ/LexUriServ.do?uri=OJ:C:2010:083:FULL:EN:PDF.
Accessed: 08.09.2011.

http://www.europarl.europa.eu/sides/getDoc.do?pubRef=-
//EP//TEXT+REPORT+A6-2007-0351+0+DOC+XML+V0//EN.    Accessed:
08.09.2011.